\documentclass[sigconf, nonacm]{acmart}

\usepackage{stfloats}
\usepackage{amsmath}
\usepackage{subcaption}
\usepackage{multirow}
\usepackage{float}
\usepackage[ruled,vlined]{algorithm2e}
\usepackage{balance}

\newcommand\vldbauthors{\authors}
\newcommand\vldbtitle{\shorttitle} 
\newcommand\vldbpagestyle{empty} 
\newcommand\vldbyear{2021}

\begin{document}
\title{Lookup or Exploratory: What is Your Search Intent?}

\author{Manoj K Agarwal}
\email{agarwalm@microsoft.com}
\affiliation{%
  \institution{Search Technology Center, Microsoft}
  \streetaddress{}
  \city{Hyderabad}
  \state{India}
}

\author{Tezan Sahu}
\email{tezansahu@microsoft.com}
\affiliation{%
  \institution{Search Technology Center, Microsoft}
  \streetaddress{}
  \city{Hyderabad}
  \state{India}
}

\begin{abstract}
Users' search query specificity is broadly divided into two categories: \emph{Exploratory} or \emph{Lookup}. If a user's query specificity can be identified at the run time, it can be used to significantly improve the search results as well as quality of suggestions to alter the query. However, with millions of queries coming every day on a commercial search engine, it is non-trivial to develop a horizontal technique to determine query specificity at run time. Existing techniques suffer either from lack of enough training data or are dependent on information such as query length or session information. In this paper, we show that such methodologies are inadequate or at times misleading. 

We propose a novel methodology, to overcome these limitations. First, we demonstrate a \emph{heuristic-based method} to identify Exploratory or Lookup intent queries at scale, classifying millions of queries into the two classes with a high accuracy, as shown in our experiments. Our methodology is not dependent on session data or on query length. Next, we train a \emph{transformer-based deep neural network} to classify the queries into one of the two classes at run time. Our method uses a bidirectional GRU initialized with pretrained \texttt{BERT-base-uncased} embeddings and an augmented triplet loss to classify the intent of queries without using any session data. We also introduce a novel \emph{Semi-Greedy Iterative Training} approach to fine-tune our model. Our model is deployable for real time query specificity identification with response time of less than $1ms$. Our technique is generic, and the results have valuable implications for improving the quality of search results and suggestions.
\end{abstract}

\maketitle

\pagestyle{\vldbpagestyle}
\begingroup\small\noindent\raggedright\textbf{Reference Format:}\\
\vldbauthors. \vldbtitle . \vldbyear.\\
\endgroup
\begingroup
\renewcommand\thefootnote{}\footnote{\noindent
This work is licensed under the Creative Commons BY-NC-ND 4.0 International License. Visit \url{https://creativecommons.org/licenses/by-nc-nd/4.0/} to view a copy of this license. 
}
\addtocounter{footnote}{-1}
\endgroup


\section{Introduction}

An important aspect of query understanding involves determining the \emph{specificity} of the query. Identifying the specificity of queries entails classifying them on a granularity spectrum of narrow to broad for a given topic \cite{Hafernik13}. The seminal work by Marchionini \cite{Marchionini06} suggests that user search activity can be divided into two broad categories: \emph{exploratory} and \emph{lookup}. Such classification of queries significantly helps understand the user search intent and hence improves the search experience. Although there are many attempts in the literature to define \textit{exploratory} or \textit{lookup} queries, there is no universally accepted definition for such categories \cite{Peters18}. In general, exploratory search queries can be viewed as information seeking, open-ended, and multifaceted \cite{White09}, while lookup queries are considered to have more narrow search goals, such as question-answering, known item search or fact retrieval, and generally have a specific answer.

Users can be in an exploratory mode due to two different reasons:
\begin{enumerate}
    \item Users do not have full knowledge or context to formulate an exact query that fulfills their information needs
    \item Users are indeed interested in learning about different facets of the topic under search
\end{enumerate}
 
In both the cases, it is likely to help users if they are presented with the search results covering different potential aspects of their search intent, that can enable them to narrow their search intent. On the other hand, if the query is a lookup query \cite{Marchionini06}, the suggestions and search results must be focused on that intent only. Hence, if the user’s query specificity can be determined at the run time, such classification can be helpful to improve the user experience substantially. Such query intent categorization systems have shown to significantly improve the user search experience \cite{Zhang19}. Various user studies show that user behaviour is different for different search specificities \cite{Athukorala14, Athukorala16}, highlighting the need to have such classification. By classifying the queries according to their specificity, we can improve the quality of the search results as well as enhance the quality of related query suggestions \cite{Sadikov10} or autosuggest \cite{Li17}, helping users complete their tasks faster. Therefore, this is a long studied problem \cite{Marchionini06, White09, Athukorala16}.


There have been studies on search intent, using specifically designed tasks assigned to a few users, in a controlled environment \cite{Lee05, Athukorala14, Athukorala16}. With millions of unique queries having different information needs appearing daily on a commercial search engine like Bing, it is difficult to model such variety in the user data using approaches involving limited data, thus either limiting the generalizability of such techniques or suffering from low accuracy.

A few systems depend on the user session information to determine the user search intent specificity \cite{Mauro18}. However, session data is highly sparse, i.e., most session features, such as query inter arrival time, user click information, and order of queries with similar information need, are difficult to replicate across sessions. Such session information requirements not only make the existing techniques impractical to determine the query specificity at run time, but we also show that some of these features are even misleading, when their distribution for queries with different specificities was studied at scale.

In this paper, we propose a novel method to classify the user search intent without using the session data. We propose a novel heuristic, using a Query-URL graph, built over search data log, to label the queries as either Lookup or Exploratory. Our methodology comprises the following steps:

\begin{enumerate}
    \item We build a Query-URL graph over the query search log. For each query $q$, we identify the related queries using the graph walk based method and by identifying the dense subgraphs of closely connected queries on this graph (Section \ref{data_prep_bipartite})
    
    \item On the set of related queries for a query $q$, identified in Step 1, we identify the top recurring patterns, denoting the top search intents for these queries (Section \ref{data_prep_phrases})
    
    \item We label the queries as Exploratory or Lookup, based on the diversity in the top recurring search patterns (Section \ref{data_prep_narrow_broad}). The basic intuition is, if the query is Lookup intent, there should not be too much semantic diversity in the top recurring patterns in the set of related queries for a given query, and vise-a-versa for Exploratory intent queries.
\end{enumerate}

Our heuristic allows us to label millions of queries into exploratory or lookup intents, covering a wide variety in the user search objectives. Ours is a first technique to label the queries into the two classes at this scale, that further enables us to study many statistical properties of such queries. We verify the quality of this data using a human judgement process over a randomly selected set of queries. We use this data to train a novel transformer-based model \cite{Vaswani17} and train it using an augmented triplet loss function, to classify the user’s search intent as lookup or exploratory, based on the query itself. For example, the query \emph{`what is normal blood oxygen level'} has lookup intent, whereas \emph{`vegetable garden'} has been labeled as exploratory intent. 


Our results show significant improvements over baseline method. Specifically, we make following contributions:

\begin{itemize}
    \item To the best of our knowledge, ours is the first method to develop a heuristic to label the queries as Exploratory or Lookup without using the session data.
    
    \item Our methodology is generic, and we enable query specificity classification at scale. Our dataset comprises over $14 M$ queries, three order of magnitude higher than any existing dataset \cite{Lee05, Athukorala14, Athukorala16}, covering a wide range of search intents. The accuracy of our method is over 81\% against a test set sampled from our dataset and judged through humans.
    
    
    \item We demonstrate that long-held insights such as query length \cite{Herrera10, Hafernik13, Athukorala14, Athukorala16, Devapujula19}, and the position of the query in a user search session \cite{White07}, to determine the query specificity, do not hold strongly by analyzing the behavior of user queries on the real search data at scale.
    
    \item We develop a transformer-based deep learning model and train it using an \emph{augmented triplet loss} to achieve an accuracy of $80 \%$ on the human-judged test set. Our results demonstrate that the performance of our model over the test data is significantly better than random labelling of the query.
    
    \item Finally, we propose a novel \emph{Semi-Greedy Iterative Training} algorithm to improve the model accuracy to $ 87 \%$ using pseudo-ground truths.

\end{itemize}


As explained in Section \ref{data_prep_bipartite}, for a given query, our heuristic needs a minimum number of related queries in the search log, to be able to classify the query intent. Hence, a large fraction of tail queries are not labeled directly by our heuristic. But we demonstrate that our transformer-based model trained using the heuristic labels on head and body queries is able to generalize and classify tail queries upto an accuracy of $\sim 74 \%$.

The rest of the paper is organized as follows: Section \ref{related_work} covers the related work.  In Section \ref{definitions}, we formally define Lookup and Exploratory queries. Subsequently, in Section \ref{data_prep}, we present our heuristic-based methodology to identify the query intent specificity. In Section \ref{model}, we describe our transformer-based model approach. In Section \ref{experimentation}, we showcase the performance of our heuristic and transformer-based models
over human-judged test data and also propose two novel iterative training algorithms to improve their performance.  Finally, in Section \ref{analysis_discussion}, we present our analysis followed by conclusion in Section \ref{conclusion}.


\section{Related Work} \label{related_work}
Study of search behavior and associated goals has been an area of keen interest. The web is extremely dynamic and unstructured \cite{Mansourian04} and it is non-trivial to analyze the user search behavior. Several studies in the past have tried to compare the user behavior for different types of information needs and investigated several trends and reasons. Attempts to study the relationship between user behavior (in terms of query formulation) and their goal has also led to the formulation and use of heuristics like \emph{Task Difficulty} \cite{Downey08}, \emph{Coherence Score} \cite{Devapujula19} and \emph{Information Gain} \cite{Athukorala14}, which can be used to identify the search goals. The study by Athukorala et. al. \cite{Athukorala16} summarizes several past investigations into the effect of search goal, difficulty, complexity, and user knowledge on information search behavior.

Query intent classification has become an integral part of search engines and plays an important role in vertical search \cite{Li08} and sponsored search \cite{Broder07}. Such classification has been researched upon previously and several studies suggest the categorization of queries into  \emph{navigational}, \emph{informational} and \emph{transactional} \cite{Broder02, Kang03, Lee05}. Marchionini suggested a slightly different categorization for search intents, i.e., \emph{exploratory} and \emph{lookup}, with the initial focus towards IR systems \cite{Marchionini06}. Research on categorizing search queries into these intents was carried out by investigating specific features such as parts of speech, query length \cite{Hafernik13}, click through rates, task completion time and scroll depth \cite{Athukorala16}.

Preliminary efforts to build classical machine-learning based classifiers for such a task \cite{Kang03, Lee05, Herrera10, Athukorala14, Athukorala16, Devapujula19} make heavy use of session-related information apart from query and URL related information to achieve decent accuracy. A summary of their techniques and results is provided in Table \ref{tab:classifier_summary}. Other approaches for classifying query intent involve the use of query-click bipartite graph \cite{Li08} and pseudo relevance feedback \cite{Shen06}. 

\begin{table*}[htbp]
  \caption{Classical ML-based query intent specificity classifiers}
  \label{tab:classifier_summary}
  \begin{tabular}{p{2.1cm} p{3.5cm} p{2.1cm} p{3.2cm} p{1.2cm} p{2.2cm}}
    \toprule
    Reference & Classification Categories & Dataset & Features & Model & Performance\\
    \midrule
    Kang et al. ('03) & informational, navigational & TREC 2000-01 & difference of distribution, usage rate as anchor texts, POS information, mutual information & regression & 91.7\% precision \& 61.5\% recall\\
    
    Lee et al. ('05) & informational, navigational & 50 most popular queries issued to Google from the UCLA CS Dept. & past user click behavior, anchor-link distribution & regression & 90\% accuracy\\
    
    Herrera et al. ('10) & navigational, informational, transactional & WT10g & query length, URL match ratio, title match ratio, terms, popularity & SVM & 79.18\% precision \& 79.18\% recall\\
    
    Athukorala et al. ('14) & broad, narrow, intermediate & prepared by senior researchers from 6 CS disciplines & no. of articles seen, no. of articles clicked & decision tree & 72.1\% accuracy \& 0.687 AUC \\
    
    Athukorala et al. ('16) & lookup, exploratory & subset of articles on arXiv & cumulative clicks, query length, maximum scroll depth & random forests & 85\% accuracy \& 0.859 AUC\\
    
    Devapujula et al. ('19) & broad, narrow & manually labelled random sample of 100k queries & query length, coherence score, word2vec encodings, number of words & SVM & 81.2\% accuracy\\
    \bottomrule
  \end{tabular}
\end{table*}

However, most of these studies involved using specifically designed tasks in a controlled environment \cite{Lee05, Athukorala14, Athukorala16}. These tasks are assigned to actual users and their search queries are analyzed to place the queries under Exploratory or Lookup intents. Naturally, such studies are limited by the types and number of tasks designed and conducted. Moreover, the use of an extremely small dataset (pertaining to a niche segment of topics) for these tasks prohibits the extension of such studies and models to the scale at which web search takes place. Further, these models use session-related information as features apart from the query and URL related features for classification of the intent. This restricts the utility of such models because they may not be able to classify queries right at the query time, which is essential for such a model to be useful in offering better search results.

\begin{figure}[!hb]
  \centering
  \includegraphics[width=1\linewidth]{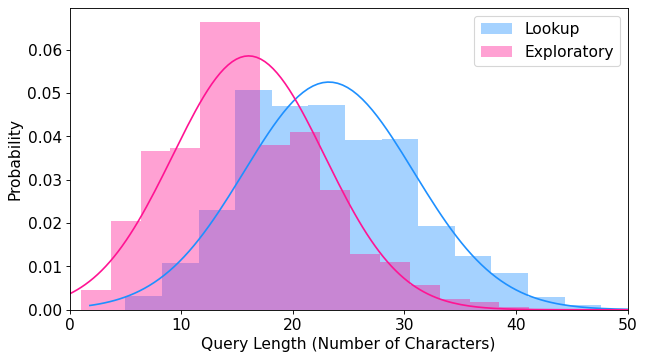}
  \caption{Query length distributions for exploratory and lookup intent queries}
  \label{fig:query_length}
\end{figure}

Among the many features used for intent classification, length of the query has been claimed to be a key differentiator \cite{Herrera10, Hafernik13, Athukorala14, Athukorala16, Devapujula19}. The work by Bendersky et al. \cite{Bendersky09} involves a deeper analysis of long queries and concludes that the click behavior (and search effectiveness) is negatively correlated with query length, primarily because lengthy queries represent complex and specific information requirements, which may be difficult to retrieve. 
In \cite{Athukorala16}, the authors report that question intent queries (Lookup) are shorter than comparison queries (Exploratory). To examine this claim, we plot the distribution of the query length belonging to the two classes based on our approach (Figure \ref{fig:query_length}). Clearly, there is a significant overlap between the two distributions, and it shows that it is infeasible to define the query specificity primarily based on the query length.

In this paper, we develop a novel heuristic using user search logs, to label the queries offline based on their specificity without using any session information. The data labeled using our heuristic is judged through a human-judgement process and shows the high accuracy of our heuristic.
Further, we use this labeled data to develop a transformer-based deep learning model to classify the query at runtime according to its specificity. We also present a new Semi-Greedy Iterative Training approach to fine-tune our model and show the boost in its performance achieved by doing so.

We derive our inspiration from the many pretrained models such as GPT-2 \cite{Radford19}, BERT \cite{Devlin19} and XLNet \cite{Yang20}, which demonstrate the ability of Transformers to perform a wide range of NLP-related tasks \cite{Wolf20}. Universal representation of queries through contextual embeddings like BERT \cite{Devlin19} and ELMo \cite{Peters18} help capture general language semantics and may show promising results in modeling the user intent effectively. The \emph{GEN Encoder} \cite{Zhang19} is one such robust system that learns a distributed representation space for user intent from user feedback in web search. Using an efficient approximate nearest neighbor search, it has also been demonstrated to reflect certain information seeking behaviors in search sessions.

\section{Defining Lookup and Exploratory} \label{definitions}

There have been many attempts in the literature to define exploratory search \cite{White09, Palagi17, Athukorala16}. In \cite{Palagi17}, the authors present the detailed survey.
With respect to a query, we capture these observations and define Lookup or Exploratory intent queries as follows:

\vspace{5pt}

\textbf{Lookup/Narrow Queries:} A lookup query $q$ is of the form $\{q \rightarrow I\ \rightarrow A\}$, where $A$ is a search result directly satisfying the user search intent. 
Such queries are aimed at \emph{fact retrieval}, \emph{question intent queries} or \emph{known item search queries}. \cite{Marchionini06}. Examples of Narrow queries include, \emph{`height of Mt. Everest'}, \emph{`how much is the high fever in children'}, \emph{`amex card late fee'}, etc.

Such queries have a specific answer and any deviation from the intent results in incorrect or partial response to the user's search objective, thus affecting the user experience adversely.

\vspace{5pt}

\textbf{Exploratory/Broader Queries:} An exploratory query $q$ is of the form $\{q \rightarrow I^* \rightarrow A^* \}$ where the query $q$ can be expanded or interpreted into one or more intents $I^*$ such that $\; \exists \; q_i \in I^* \; | \; q_i \rightarrow A^*, \; |A^*| > 1$, i.e., given a query $q$, there can be one or more interpretations and at least one of these interpretation $q_i$ has multiple responses or perspectives. 

In other words, lookup/narrow intent and exploratory/broader intent query classes are disjoint.

Broader queries can be interpreted in multiple ways. For instance, the query \emph{`vegetable garden'} can be interpreted by a search engine to mean \emph{`how to set up a vegetable garden'} or \emph{`kitchen vegetable garden'} or \emph{`cost of setting up a vegetable garden'}. A commercial search engine may present a mix of these interpretations in its results or show more generic URLs serving multiple intents. Further, results for a given interpretation can be subjective in nature and can potentially have multiple answers. Such queries usually emerge at the start of a complex task, knowledge acquisition or learning process \cite{Downey08}.

As mentioned in Section \ref{related_work}, most existing approaches in the literature depend on the session data to determine query specificity to determine if a query is Broader or Narrow \cite{Mauro18}. Many of these approaches assume that for a complex search task, users typically start with a broader or exploratory intent query and progressively narrow their queries until they reach their goal \cite{Downey08}. However, such assumptions have multiple shortcomings. First, a complex search task for a user may be distributed over multiple search sessions. In such cases, even the initial queries in a new search session can be continuation of a previous search session with the same task and may start with specific intent queries, building upon the knowledge acquired in the previous sessions. We demonstrate (Section \ref{expt2}), using the search logs over a commercial search engine, that unlike the hitherto established insight \cite{White07}, the specificity of a query has no correlation with its position in a user session. Second, a user may already be an expert user and may directly start his search task with specific intent queries. Third, a typical user session is intermingled with different search tasks. If a user is searching more than one topic simultaneously, is it non trivial to partition the user session into a cohesive sequence of queries serving a specific search intent \cite{Zhang19}. Finally, session data is highly sparse. Very few sessions follow the same sequence of queries, even if they share the same search goal. Most of the search sessions 
may have never appeared in the past. Therefore, a session-based approach generally relies on the features such as \emph{position of a query} in a session, \emph{inter query time} and \emph{length of a session} \cite{Hienert18, Downey08}, making such approaches impractical in a real time environment. 

Contrary to existing approaches, in this paper, we propose a novel approach, which does not rely on session characteristics to determine the query intent. 

\section{Heuristic-Based Classification} \label{data_prep}

In this section, we present our heuristic to identify \textit{exploratory} and \textit{lookup} intent queries as defined in the previous section. Towards this goal, for a given query, we first identify the set of its related queries using Query-URL graph (Section \ref{data_prep_bipartite}). Next, we identify the top search intents in this set of related queries (Section \ref{data_prep_phrases}). In Section \ref{data_prep_narrow_broad}, we identify a query as \textit{exploratory} or \textit{lookup} based on the diversity in these search intents.

\subsection{Query-URL Bipartite Graph} \label{data_prep_bipartite}

\begin{figure}[!ht]
  \centering
  \includegraphics[width=\linewidth]{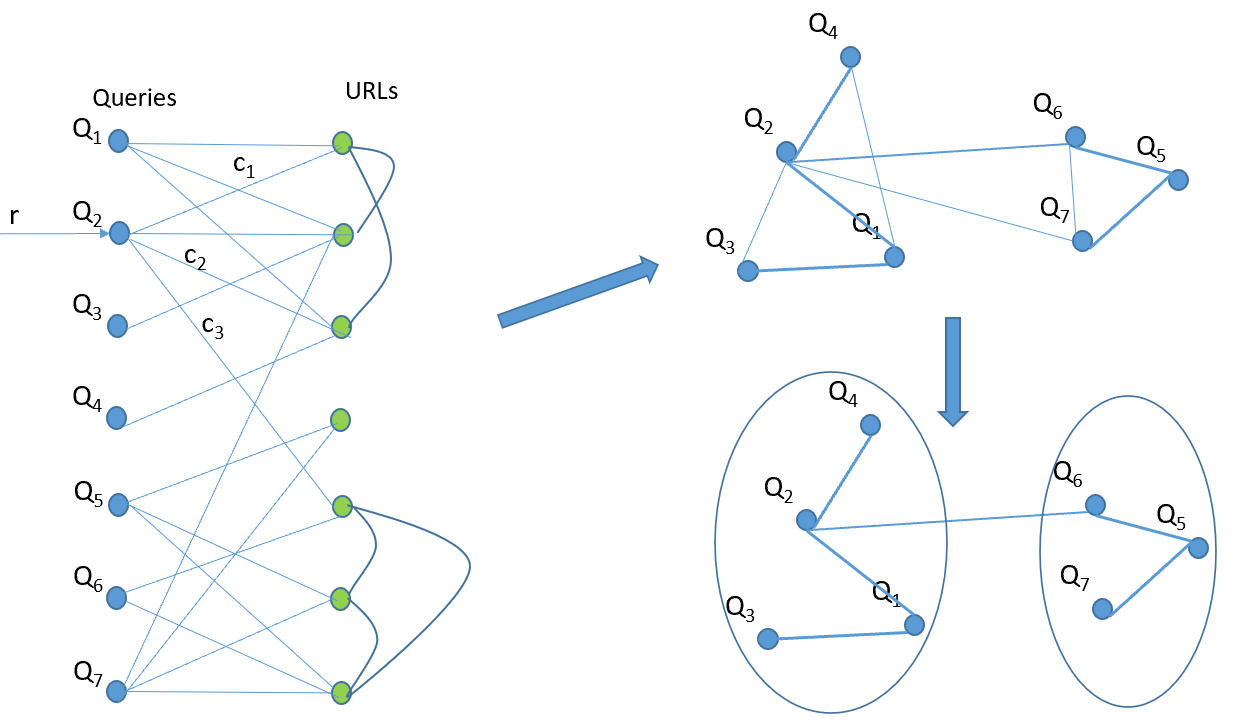}
  \caption{Induction of Query-URL Bipartite Graph and Query-Query Graph on search traffic}
  \label{fig:query_url}
\end{figure}

Search log can be represented as a Query-URL bipartite graph, as shown in Figure \ref{fig:query_url}. We use two different algorithms to collect related queries using the search log. In the first algorithm, we use \emph{Random Walk} \cite{Craswell07} over the graph to compute the hitting time between different queries. Starting from query $q_i$ in the Query-URL graph, the expected time taken to reach query $q_j$ in the random walk over the Query-URL graph is called the \emph{hitting time} \cite{Mei08}. Smaller the hitting time, the closer are the two queries. The intuition is, if two queries share many URLs (i.e., semantically closer to each other in the Query-URL graph), their hitting time must be smaller. For a given query $q$, all queries with their hitting time below a specified threshold are considered related. For each query in the Query-URL graph, the algorithm generates an ordered list of related queries, in ascending order of their hitting times. 

In our second algorithm \cite{Agarwal20}, we first induce a Query-Query graph ($QQ$ graph) from the Query-URL graph (Figure \ref{fig:query_url}). The weights between the queries are computed based on the number of URLs they share between them in the Query-URL graph. More the URLs they share, more is the edge weight between them in the $QQ$ graph. As explained in \cite{Agarwal20}, the edge weight $0 < w_e \leq 1$ in the $QQ$ graph. On this induced graph, query clusters are identified such that each cluster has the target \emph{goodness score}, where goodness of a cluster is the function of its density and the weight of the edges in the cluster. If a cluster reaches the goodness above the target score, it is accepted, or else it is further partitioned. At the end of this process, the queries in the $QQ$ graphs are partitioned into clusters where queries within a cluster are semantically correlated. The cluster sizes may vary, depending on the actual number of queries in the search log which share the same intent. Finally, for a cluster $C$, the algorithm produces a pairwise relation score for each query pair $q_i, \; q_j \in C$, computed as follows:

\begin{equation}
    R_{ij} = argmax_{\phi \in \Phi(i, j)} \prod_{e \in \phi} w_e
\end{equation}

where $\phi$ is the path between $q_i$ and $q_j$ in $QQ$ graph, and $\Phi(i,j)$ is set of all the paths between $q_i$, $q_j$ in the partitioned graph induced over queries in cluster $C$. For a query $q$ in a cluster $C$, the other queries with relation score above a threshold are considered its related queries. 

For a given query, we take a union of related queries sets generated by the two algorithms to create a final set of related queries. Once these are identified, we consider only those queries, which contain at least $n = 40$ related queries. The choice of \textit{n} is explained in Section 4.3. Though many related queries identified by both the algorithms are common in the two sets, the first algorithm may provide queries with relatively diverse intents (breath), whereas queries identified by second algorithm are more semantically correlated (depth).

\subsection{Identifying Patterns in the Query Cluster} \label{data_prep_phrases}

For a given query $q$, let set of queries $Q_q = \{q_1, q_2, ..., q_n\}$ be the set of its related queries.  

We analyze the queries in set $Q_q$ to identify the top implicit or explicit intent of query $q$. Users can use different words to express the same intents, therefore, we first identify the top recurring patterns in the set $Q_q$, to handle the variations in the query formulations by different users, expressing same search intent. The recurring patterns across the queries in set $Q_q$ represent the top intents. Since the word order matters in specifying the intent, patterns are an ordered sequence of words. For instance, for a query $q = $ \emph{`virat kohli odi average'}, let its set of related queries $Q_q$ contain a frequent pattern $p = $ \emph{`virat odi'}. This pattern may be occurring in a subset $Q_p \subseteq Q_q$, of queries. $\forall q \in Q_p$, the words \emph{virat} and \emph{odi} appear in the same order. Each pattern is true with respect to a subset of queries in set $Q_q$. Thus, we create a dictionary $P_q$ of patterns in the form of:

$$p \rightarrow Q_p | \; \forall p \in P_q, \; \exists \;  Q_p \subseteq Q_q \; s.t. \; | Q_p | \geq k$$

\vspace{10pt}
Each pattern must appear in at least $\delta \cdot n \geq k$ queries. By modifying $\delta$ or $k$ we can control the number of patterns discovered. For two patterns, $p_i$ and $p_j$, if $|Q_{p_i} \cap  Q_{p_j}|\geq \gamma \cdot m$, where $ m \geq min(|Q_{p_i}|, |Q_{p_j}|) $, they are merged into a single pattern. $\delta$ must be set to a low value and $\gamma$ must be set to a high value such that $0 < \delta < \gamma \leq 1$. In our experiments, we found that $\delta = 0.1$ and $\gamma = 0.8$ worked well. In Algorithm \ref{alg:phrase_id}, we present the pseudo code to identify the patterns in the Query set.

\begin{algorithm}
\SetAlgoLined
     $\boldsymbol{N} \gets$ getFreqKeywords($Q_q$, $k$)\;
     \tcc{Set of keywords with frequency above $k$ in related query set $Q_q$}
     \
     
     $\boldsymbol{T_{set}} \gets$ initRules($N$)\;
     \tcc{Set of discovered patterns over input query set $Q_q$, initialized with frequent words}
     \
     
     $\boldsymbol{T_{mod}} \gets T_{set}$\;
     \
     
     \While{$T_{mod}$.size > 0}{
        $T_{mod}$.clear()\;
        $T_{changed}$.clear()\;
        \ForEach{Rule $r_i$ in $T_{set}$}{
            \ForEach{Rule $r_j$ in $T_{set}$}{
                \If{$P(r_i | r_j) > \gamma$}{
                    $T_{new} \gets$ mergePatterns($r_i$, $r_j$)\;
                    \tcc{Create ordered sequence}
                    \
                    
                    $T_{mod} \gets T_{mod} \cup T_{new}$\;
                    $T_{changed} \gets T_{changed} \cup r_i \cup r_j$\;
                }
                \ElseIf{$P(r_i | r_j) > \delta$}{
                    \If{$P(r_j | r_i) > \gamma$}{
                        $T_{new} \gets$ mergePatterns($r_j$, $r_i$)\;
                        \tcc{Create ordered sequence}
                        \
                        
                        $T_{mod} \gets T_{mod} \cup T_{new}$\;
                        $T_{changed} \gets T_{changed} \cup r_i \cup r_j$\;
                    }
                }
            }
            
            \ForEach{Rule $r$ in $T_{changed}$}{
                $T_{set} \gets T_{set} - r$\;
            }
            
            \ForEach{Rule $r$ in $T_{mod}$}{
                $T_{set} \gets T_{set} \cup r$\;
            }
        }
     }
     
    \caption{Pattern Identification Algorithm}
    \label{alg:phrase_id}
\end{algorithm}

For each pattern $p \in P_q$, discovered over a query set $Q_p \subseteq Q_q$ we assign it a weight $p_w$ as follows: 

\begin{equation}
    p_s = log(1 + f) \cdot \frac{s}{1 - c + \epsilon}
    \label{eqn:2}
\end{equation}

where $s$ is the \emph{support} of the pattern $p$, $c$ is its \emph{confidence}, and $f$ is the \emph{cumulative frequency} of queries in set $Q_p$, computed respectively as follows:

$$ s = \frac{| Q_p |}{| Q_q |}$$

$$ c = \frac{| Q_p |}{| Q_{p'} |}; \; Q_p \subseteq Q_{p'} \subseteq Q_q$$

For the pattern $p$ and $\forall q \in Q_p$, the `order' relationship is enforced, i.e., $\forall w_i, w_j \in p, q \; | \; w_i \rightarrow w_j \; in \; p \Rightarrow w_i \rightarrow w_j \; in \; q$, where $\rightarrow$ denotes \emph{`precedes'} relation. Whereas, $\forall q \in (Q_{p'} - Q_p)$ , $\exists w_i, w_j \in p, q \; | \; w_i \rightarrow w_j \; in \; p, \; but \; w_j \rightarrow w_i \; in \; q$. Therefore, for a pattern $p$, the same order of words in $p$ is not followed by the queries in set $(Q_{p'}-Q_p)$. Among the queries containing all the words in the pattern $p$, $c$ underlines the fraction of queries which follow the same word order, underlying the confidence in the discovered sequence. We accept only those patterns where $c > 0.5$. Thus, for any permutation of words, only one of the sequence can be admitted as pattern, i.e., if \emph{`virat odi'} is discovered as pattern, \emph{`odi virat'} cannot be another pattern. 

Finally, 
$$f = \sum_{q \in Q_p} f_q$$

$f_q$ is the frequency of query $q \in Q_p$ in the search log. 

Thus, the pattern weight $p_w$ is high if proportion of queries in set $Q_q$ in which it occurs is high, cumulative frequency $f$ of these queries is high, and its confidence score $c$ is high, capturing its overall popularity in the query set $Q_q$, w.r.t. query characteristics in the set $Q_q$. Further, $\forall q \in (Q_q - Q_{p'}), \; \exists w \; | \; w \in p, \; w \notin q$. 

\subsection{Exploratory or Lookup} \label{data_prep_narrow_broad}

For Lookup intent queries, the top URLs serve the specific query intent. Hence, the queries in set $Q_q$ for such queries are more likely to be semantically similar (as such URLs will come as top search results only for similar intent queries). Conversely, for Exploratory intent queries, there will be more diversity in the query intents in the set $Q_q$ (as such URLs are likely to serve multiple intents in broader intent queries). In other words, the patterns discovered over set $Q_q$ are likely to be more diversified for \textit{Exploratory} intent queries and vise-a-versa for \textit{Lookup} intent queries. This observation is captured by analyzing the graph structure induced over the discovered pattern. 

Patterns represent the top search intents for the queries in the set of related queries $Q_q$. Our objective is to determine the diversity in these search intents. For this, we divide the patterns into semantically similar set of patterns. We use the \texttt{BERT-base-uncased} encodings \cite{Turc19}, and compute the pairwise cosine similarity between the discovered patterns. We induce a weighted graph $G (V, E)$ between the patterns where where a node $v\in V$ in the graph corresponds to a pattern $p \in P_q$ discovered over query set $Q_q$ and an edge is induced between two patterns if their cosine similarity is above a threshold of $0.8$. Each node is assigned a weight corresponding to its pattern weight as per Eq \ref{eqn:2}. 


To identify the structures embedded in the graph with a  high confidence, we ensure that the related query sets for the queries are big enough. Therefore, $40 \leq | Q_q | \leq 1000$. We ensure that each set contains at least 40 queries in set $| Q_q |$ for any query \textit{q}. The patterns discovered over small size query sets will not be diverse by definition. Thus, the queries with less than 40 related queries (tail queries) are not considered in preparing our training data. Similarly, if a set contains too many related queries, we consider only the most closely related 1000 queries in set $| Q_q |$, else the computation becomes prohibitively expensive for a massive dataset comprising millions of large sets. Note, the number of large sets grow linearly in terms of number of related queries, i.e., if there is a query with $n = 10000$ related queries, there will be $O(n)$ sets of size 10000 each (the relation between queries $q_i$ and $q_j$ in the Query-URL graph is symmetric). Thus the overall computation increases by $O(n^2)$. By restricting the maximum size of $n$ to 1000 closely related queries, we ensure the completion of the computation within an acceptable time limit, while simultaneously capturing the top search intents.

\subsection{Dense Graph} \label{data_prep_dense_graph}

The graph density is captured as follows: The induced graph $G (V, E)$ can be disconnected. We first identify the biggest connected component $G'(V',E') \; | \; V' \subseteq V, \; E' \subseteq E$, where the graph size is determined based on the number of nodes in the graph (not on the node weight). We identify the \emph{k-core} in $G'$ as the dense graph. In our experiments, we have set $k=2$. For an identified $k$-core graph $G_k$, we compute its weight, $w_k(G')$ as follows:
\begin{equation}
    w_k(G') = \frac{\sum_{v \in V'} w_v}{W}
\end{equation}

where, $w_v$ is the weight of a node, computed as per Eq \ref{eqn:2} and W is computed as follows:
\begin{equation}
    W = \sum_{v \in V} w_v
\end{equation}

\begin{figure}[!ht]
    \centering
    \begin{subfigure}[b]{0.45\linewidth}
        \includegraphics[width=\linewidth]{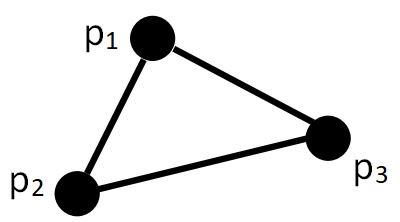}
        \caption{2-core graph}
        \label{fig:2core}
    \end{subfigure}
    \hfill
    \begin{subfigure}[b]{0.45\linewidth}
        \includegraphics[width=\linewidth]{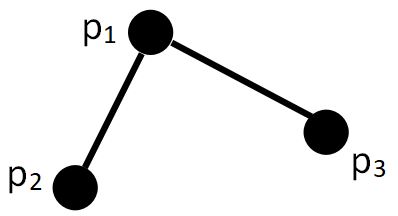}
        \caption{Not a 2-core graph}
        \label{fig:not_2core}
    \end{subfigure}
    
    \caption{Example of non $k$-core and $k$-core graphs with $k = 2$}
    \label{fig:k-core_examples}
\end{figure}


Fig. \ref{fig:2core} and \ref{fig:not_2core} depict graphs induced over three patterns $p_1$, $p_2$ and $p_3$ that are 2-core and not 2-core respectively.

If $w_k(G')$ is above a high threshold $T_h$, it indicates, the patterns representing the top intents over set $Q_q$ are densely connected, thus semantically similar, indicating the query $q$ has a specific search goal and is lookup intent. On the other hand, if the weight $w_k(G')$ is below a low threshold $T_l$, we consider the top intents are too diverse. In our experiments, we have set $T_h = 0.9$ and $T_l = 0.5$. The threshold $T_h$, and $T_l$ are set conservatively, to ensure the quality of the queries marked as exploratory or lookup by our heuristic is good. Thus we ignored those queries which our heuristic could not mark with high confidence. Note that these queries do not constitute any separate class of relatively \emph{hard} queries from the specificity point of view. It is just that the resulting structure of their underlying Query-URL graph in the search logs was inadequate to capture their specificity for our heuristic. Hence, these queries are not considered for training our model, explained in Section \ref{model}.


\subsection{Dataset Preparation} \label{data_dataset_prep}

We obtained these intent labels for $25M$ queries from Bing
logs, out of which $11M$ are marked as Lookup intent, 
$3.1M$ marked as Exploratory intent and the remaining were marked as Ambiguous intent by the heuristic. The exploratory and lookup queries, along with their
suggestions were organized into triplets of anchor, positive and negative, to be used to compute the triplet loss, which is discussed in Section \ref{model_loss}. We further removed those triplets from the data which were obviously contradictory. 
We used query embeddings
trained over Bing search logs, to determine the semantic similarity between the queries. The contradictory triplets, i.e., those queries with high semantic similarity computed based on the cosine distance between their AGI vectors but with different specificity labels are removed, to further clean the training data. For the actual training of our classifier, we downsampled the data to contain $32000$ triplets.

\section{Transformer-Based Model} \label{model}

In this section, we present our objective function, our approach for developing the model architecture and the details related to training the model. We use \texttt{PyTorch 1.6.0} to implement our model, and train it on an NVIDIA T4, 16 GB GPU.

\begin{figure*}[ht!]
  \centering
  \includegraphics[width=0.9\linewidth]{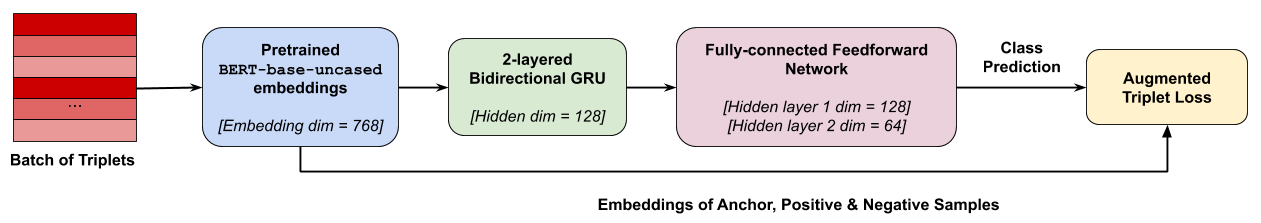}
  \caption{The architecture for our query specificity classification model}
  \label{fig:model_architecture}
\end{figure*}

\subsection{Augmented Triplet Loss Function} \label{model_loss}

Embeddings serve as an intuitive way to represent queries in a generalized fashion and model user intent in a latent space. Executing the task of query intent specificity classification successfully requires us to not only classify queries into exploratory or lookup based on the query embedding, but also separate similar looking queries (mostly offered by search engines as \emph{related searches}) having opposite intent specificities, further apart in the latent space. As an example, the queries \emph{`cake'} (exploratory) and \emph{`top 10 cake recipies'} (lookup) may both appear as related search suggestions for the query \emph{`dessert options'} (exploratory), but their representations in the embedding space must be adequately differentiated by our classifier. This motivates us to optimize a function similar to the \emph{Triplet Margin Loss}, which has demonstrated significant improvements in modeling contextual visual similarity \cite{Wang16} and image information retrieval tasks \cite{Hoffer18}.

Learning with triplets \cite{Balntas16} involves training from tuples of the form \textbf{\emph{\{a, p, n\}}}, where \textbf{\emph{a}} is referred to the \emph{anchor}, \textbf{\emph{p}} is known as \emph{positive} sample (a different sample of the same class as \textbf{\emph{a}}), while \textbf{\emph{n}} is called \emph{negative} (a sample belonging to a different class compared to \textbf{\emph{a}}). The Triplet Margin Loss minimizes the distance between \textbf{\emph{a}} and \textbf{\emph{p}}, and maximizes the distance between \textbf{\emph{a}} and \textbf{\emph{n}}, while trying to maintain an arbitrary margin $\mu$.

In the context of our work, we aim to classify query intents while also trying to differentiate the intents of related suggestions for a query based on their intent specificities. Hence, our query forms the anchor, a suggestion with similar specificity label forms the positive instance, while a suggestion with opposite label forms the negative instance. We use the Euclidean norm as the distance metric between queries in the embedding space.

Moreover, since the core objective here is a binary classification problem, it is only reasonable that we incorporate the \emph{Binary Cross-Entropy (BCE) Loss} between the anchor prediction $\hat{a_l}$ and the actual label $a_l$ as a component of the overall loss. 

Taking inspiration from multiobjective optimization \cite{Chankong83}, we formulate the objective function to be a combination of the objectives of intent classification and separation of dissimilar intents in the embedding space. To balance the contributions of each of these components to the overall loss, we introduce an arbitrary weighting factor $\eta$ as a hyperparameter to specify the relative importance of the Triplet Loss wrt the BCE Loss. Thus, the overall objective to be minimized can be expressed as follows:

\begin{equation}
    \begin{split}
        Loss(\boldsymbol{a}, \boldsymbol{p}, \boldsymbol{n}, \hat{a_l}, a_l) = \eta \cdot max\{d(\boldsymbol{a}, \boldsymbol{p}) - d(\boldsymbol{a}, \boldsymbol{n}) + \mu, 0\} \; + \\ 
        (1 - \eta) \cdot [-a_l \; log(\hat{a_l}) - (1 - a_l) \; log(1 - \hat{a_l})] 
    \end{split}
\end{equation}

where $d(\boldsymbol{x}, \boldsymbol{y}) = || \boldsymbol{x} - \boldsymbol{y} ||^2_2$  \hfill \textit{[Euclidean Distance]}

\subsection{Model Architecture and Learning}

Each sample in the batch is a triplet, containing an anchor query, a positive and a negative suggestion, is tokenized and fed into our classifier. In order to represent individual queries, we insert an external [CLS] tokens at the start of each query, which when embedded, serve as a contextualized encoding for the associated word sequence after it \cite{Liu19}. We adopt a simplified architecture for our classifier model comprising a batch input of triplets, which is passed through the pretrained 12-layer \texttt{BERT-base-uncased} embedding \cite{Turc19} (embedding dimensions $= 768$) to obtain the query representations in a latent space. Now, we pass the [CLS] representations of the anchor queries in our batch through a 2-layer bidirectional GRU with a hidden dimension of $128$. The final hidden layer output from the GRU is flattened and passed into a fully connected feed-forward network with 2 hidden layers of $128$ and $64$ neurons respectively to predict the binary class label. The overall  architecture is shown in Figure \ref{fig:model_architecture}. With $110.5 M+$ parameters, this model weighs in at just over 1.3 GB on disk.

We train our model on the dataset of $32000$ triplets generated as mentioned in Section \ref{data_dataset_prep} using the hyperparameters mentioned in Table \ref{tab:train_hyperparams}. We arrived at these optimal hyperparameters to train our model through offline experimental analysis. Dropout has been implemented at each layer to act as a regularizer for the network. The BERT embeddings along with classifier scores for the anchor, positive and negative are used to calculate the augmented triplet loss defined in Section \ref{model_loss}. The class labels are predicted by the $argmax(.)$ of the class scores. We use the \emph{AdamW} optimizer with a weight decay of $1 \times 10^{-2}$ to train this model for achieving lower losses and better generalization.

\begin{table}[!h]
  \caption{Hyperparameters for model training}
  \label{tab:train_hyperparams}
  \begin{tabular}{cc}
    \toprule
    Hyperparameter & Value\\
    \midrule
    Batch Size & 64\\
    Learning Rate & $5 \times 10^{-5}$\\
    Dropout & 0.1\\
    Triplet Loss Margin ($\mu$) & 0.01\\
    Triplet Loss Weightage ($\eta$) & 0.5\\
    \bottomrule
  \end{tabular}
\end{table}

After training for 1 epoch, our model achieves an accuracy of $80.46 \%$ and an F1 score of $0.7915$. This is significantly better compared to a baseline random binary classifier with an accuracy of $50 \%$ and F1 score of $0.5$. Later, in our experiments mentioned in Section \ref{experimentation}, we show that we are able to improve the accuracy and F1 score significantly using a novel iterative training approach. Moreover, the average specificity label prediction time for a query is computed to be $\sim 0.8 ms$, which is negligible compared to the average query execution time in commercial search engines. Table \ref{tab:examples} illustrates some of the examples that were correctly classified by our model along with some that were misclassified.

\begin{table*}[!hb]
    \caption{Some examples that were correctly classified and misclassified by our model}
    \label{tab:examples}
    \begin{tabular}{l|l|c|c|}
    \multicolumn{2}{c}{}&\multicolumn{2}{c}{\textbf{Predicted Label}}\\
    \cline{3-4}
    \multicolumn{2}{c|}{}& \textbf{Exploratory} & \textbf{Lookup}\\
    \cline{2-4}
    \multirow{4}{*}{\textbf{True Label}} & \multirow{2}{*}{\textbf{Exploratory}} & gas prices & job today usa\\
    & & vegetable garden & gpu cuda\\
    \cline{2-4}
    & \multirow{2}{*}{\textbf{Lookup}} & medications linked to alzheimer's & conversion ounce to gram calculator\\
    & & what is normal blood oxygen level & london broil recipies in slow cooker\\
    \cline{2-4}
    \end{tabular}
\end{table*}

\section{Experiments} \label{experimentation}


We conduct several experiments to validate and improve our classification approaches mentioned in Sections \ref{data_prep}
 and \ref{model}. Firstly, we describe the judging process used to curate our test dataset for all the experiments that follow. Then, we evaluate our heuristic-based classification approach against these judged ground truth labels. We then try to improve the performance of our model using an iterative training approach 
. Next, we study the impact of the size of the training dataset on our model. Subsequently, we demonstrate the \emph{Semi-Greedy Iterative Training (SGIT)} as a novel method to achieve much better performance by leveraging the goodness of the most recent model along with the best model during the iterative process. Finally, we evaluate the capability of our model (which has been trained on a set of non-tail queries) to generalize on a set of human-judged tail queries.

\subsection{Test Dataset and Evaluation Metrics} \label{judging}

We uniformly sample the queries of two types from our pool of $14.1 M$ queries, with probabilities so as to create a balanced judgement set. These queries are presented independently to two human judges, selected blindly from a common judge pool who are trained in this task. The judges are not shown the query labels based on our heuristics, and are asked to mark each query on a 3-point scale, for \emph{exploratory}, \emph{lookup} and \emph{not sure} respectively. Only those judgements, marked as exploratory or lookup and show consensus between the two judges are accepted. Note that this mechanism is a stricter selection criterion for identifying the human-judged labels compared to the best of three mechanism. 
Our final judged set contained $302$ representative queries, with 161 lookup and 141 exploratory intent queries. We use the overall accuracy along with the F1 score to evaluate the performance of our methods on the above mentioned human-judged test dataset.

\subsection{Validation of the Heuristic} \label{expt1}

Once the labels are obtained from the human judges, we compare them with the labels assigned based on our heuristic. 
In Table \ref{tab:expt_1_heuristic_cf}, we present the confusion matrix based on these judgements. We see that the accuracy of our heuristic is at 81\%, as well as the F1 score is 0.81. These results show that our heuristic can identify the query specificity with high accuracy. 

\begin{table}[!h]
    \caption{Confusion Matrix for Heuristic-based classification}
    \label{tab:expt_1_heuristic_cf}
    \begin{tabular}{l|l|c|c|}
    \multicolumn{2}{c}{}&\multicolumn{2}{c}{\textbf{Predicted Label}}\\
    \cline{3-4}
    \multicolumn{2}{c|}{}& \textbf{Exploratory} & \textbf{Lookup}\\
    \cline{2-4}
    \multirow{2}{*}{\textbf{True Label}}& \textbf{Exploratory} & $39.07 \%$ & $7.62 \%$\\
    \cline{2-4}
    & \textbf{Lookup} & $11.26 \%$ & $42.05 \%$\\
    \cline{2-4}
    \end{tabular}
\end{table}

\begin{figure}[!h]
  \centering
  \includegraphics[width=0.9\linewidth]{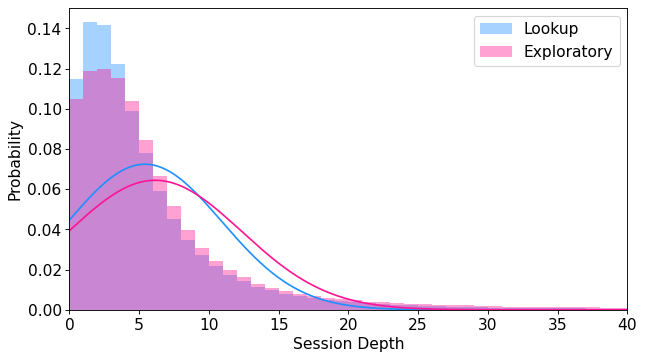}
  \caption{Distributions of query position in the session for lookup and exploratory queries in the original dataset}
  \label{fig:expt_1_session_depth}
\end{figure}

\textbf{Relation with Query Position in a Session:} In this part of the experiment, we study the correlation between the depth of the session with respect to the query specificity. In our heuristic, we did not use the session features to identify the query intent, as discussed in Section \ref{data_prep}. As shown in Figure \ref{fig:expt_1_session_depth}, there is not much discriminatory behavior between the two different type of queries based on their position in a user session. In fact, we see that lookup queries have their mode occurring before the exploratory queries. This PDF for the two types of queries is plotted over the $14.1 M$ queries marked based on our heuristic, and hence it can be considered as a statistically more representative user behavior as opposed to studies conducted in a controlled environment over limited user sessions. To our mind, this is an important insight, which can impact the way user experiments are designed for such studies.

\vspace{5pt}

\begin{figure}[!h]
  \centering
  \includegraphics[width=0.9\linewidth]{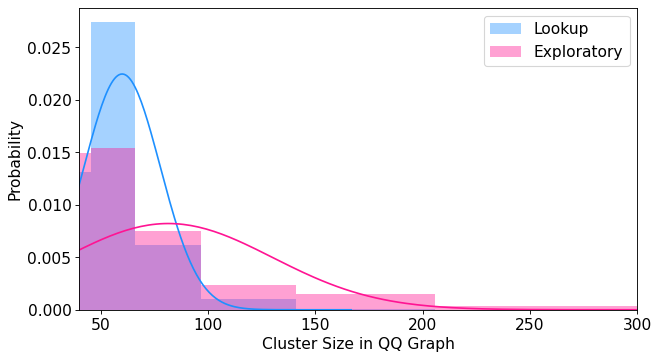}
  \caption{Distributions of related query cluster sizes in the $QQ$ Graph for lookup and exploratory intent queries in the human-judged test dataset}
  \label{fig:expt_1_related_distrib}
\end{figure}

\textbf{Relation with Related Queries:} In this part of the experiment, we verify the basic intuition behind our heuristic with respect to the human-judged test set. As stated in Section \ref{data_prep_bipartite}, the minimum related query set size for each judged query in the test set is $| \; Q_n \;|=40$. In Figure \ref{fig:expt_1_related_distrib}, we plot the related query set size histogram for queries marked as exploratory and lookup by the human judges. We observe the for the queries marked as lookup, the related query set is relatively smaller compared to queries marked as exploratory by the judges, confirming our premise that the URLs shown for exploratory queries are more diverse, and hence, such queries have relatively larger number of related query sets based on the Query-URL graph, as well as more diverse intent queries appear in the related query set for such queries. 

\subsection{Iterative Model Training} \label{expt2}

From our results in the previous experiment, it is evident that our basic transformer-based classifier is almost as good as the heuristic-base approach. This suggests that the predictions of our transformer-based classifier can potentially be used as pseudo-ground truths for further training of our model. 

To ensure the robustness of our model to noisy labels in the training data which may have crept in due to the errors made by our heuristic, we follow an iterative approach to train our model. We draw inspiration from the \emph{Iterative Trimmed Loss Minimization} approach \cite{Shen19} and the use of \emph{pseudo-ground truths} in the iterative refinement of RefineLoc \cite{Alwassel19} to develop and evaluate an iterative technique to refine our classifier.

We train our model iteratively for $T$ iterations in such a way that for each iteration a fraction $\alpha \in (0, \; 1)$ of the labeled training dataset $S$ (of total size $n$) is sampled and the classifier is trained on the $\lfloor \alpha n \rfloor$ sampled triplets. Now, the predictions of the model trained in iteration $t$ on the entire training dataset are used as pseudo-ground truths (labels) for training during iteration $i+1$. The details about this approach are described in Algorithm \ref{alg:iter_train}
.

\begin{algorithm}
\SetAlgoLined
     \SetKwInOut{Input}{input}
     \SetKwInOut{Output}{output}
     \Input{samples $S = \{(x_i, \; y_i)\}_{i=1}^n$, number of iterations $T$, fraction of samples $\alpha$}
     $\theta_0 \gets$ pretrained \texttt{BERT} embeddings $+$ random weights\;
     $\theta^* \gets \theta_0$\;
     $acc_{best} \gets accuracy(\{y_i\}_{i=1}^n, \;  \{\theta_0(x_i)\}_{i=1}^n)$\;
     \For{t = 0, 1, ..., T-1}{
        $S_t \stackrel{i.i.d}{\sim} Uniform(S) \;$ s.t. $| \; S_t \; | = \lfloor \alpha n \rfloor $\;
        $\theta_{t+1} \gets TrainClassifier(\theta_t, S_t)$\;
        \
        
        $pred \gets \{\theta_{t+1}(x_i)\}_{i=1}^n$\;
        $acc_{curr} \gets accuracy(\{y_i\}_{i=1}^n, \;  pred)$\;
        \If{$acc_{curr} > acc_{best}$}{
            $acc_{best} \gets acc_{curr}$\;
            $\theta^* \gets \theta_{t+1}$\;
        }
        
        $S \gets \{(x_i, \; pred_i)\}_{i=1}^n$\;
        
     }
     \Output{$\theta^*$}
     
    \caption{Iterative Model Training Algorithm}
    \label{alg:iter_train}
\end{algorithm}

\begin{figure}[!h]
  \centering
  \includegraphics[width=\linewidth]{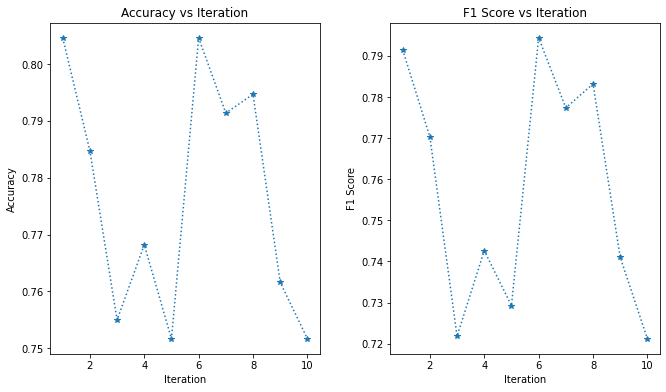}
  \caption{Change in accuracy and F1 scores of the model with each iteration}
  \label{fig:expt_2_metrics}
\end{figure}

We use this algorithm to train our model for 10 iterations by setting $\alpha = 0.8$, on the dataset containing $32000$ triplets. Notice that one can always use cross-validation to find the best $\alpha$. The performance of this model is plotted in Figure \ref{fig:expt_2_metrics}. We achieve the best performance after iteration 6, with an accuracy of $80.46 \%$ and F1 score of $0.7944$ on the judged test set. This model is able to correctly classify queries like \emph{`air conditioning not working'} (lookup) and \emph{`car rental agreement'} (exploratory), which were earlier misclassified by the model in iteration 1.

\subsection{Impact of Training Dataset Size} \label{expt3}

To assess the impact of the size of the training dataset on model performance, we create two new datasets by uniformly sampling $48000$ and $64000$ triplets respectively from the original pool of $14.1M$ triplets. Subsequently, we train our transformer-based classifier on these datasets using the iterative algorithm mentioned in Section \ref{expt2} and compare the results to those obtained on our model trained on $32000$ triplets.

\begin{figure}[!h]
  \centering
  \includegraphics[width=\linewidth]{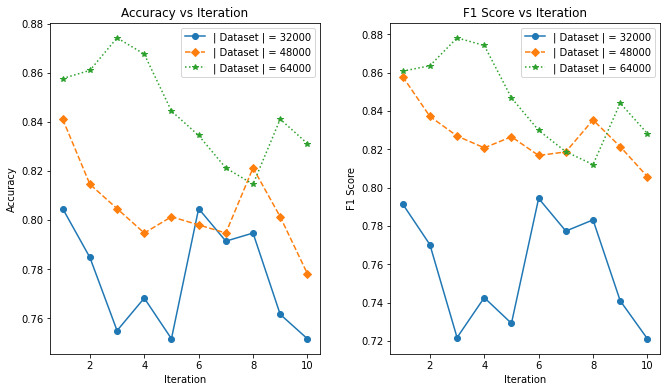}
  \caption{Accuracy and F1 scores of the models for training datasets of sizes 32000, 48000 and 64000 query triplets}
  \label{fig:expt_3_metrics}
\end{figure}

Results for this experiment can be found in Figure \ref{fig:expt_3_metrics}. It is evident from the plots that increasing the dataset assists in improving the classifier significantly. The dataset with $64000$ triplets yields the best results on our human-judged test set. The classifier performances for the 3 datasets are summarized in Table \ref{tab:expt_3_summary}. Analysing the confusion matrices suggests that the best model trained using $64000$ triplets improves its accuracy and F1 score by correcting most of the lookup intent queries like \emph{`pinnacle bank tx login'} and \emph{`what does jesus say about anger'}, which were misclassified by the best model trained only on $32000$ triplets.

\begin{table}[!h]
  \caption{Summary of model performances with different training dataset sizes}
  \label{tab:expt_3_summary}
  \begin{tabular}{ccc}
    \toprule
     Dataset Size & Best Model Accuracy & Best Model F1\\
    \midrule
    32000 & $80.46 \%$ & $0.7944$\\
    48000 & $84.11 \%$ & $0.8579$\\
    64000 & \textbf{87.41\%} & \textbf{0.8782}\\
    \bottomrule
  \end{tabular}
\end{table}

\subsection{Semi-Greedy Iterative Training (SGIT)} \label{expt4}
The iterative training algorithm mentioned in Section \ref{expt2} completely replaces the heuristic ground truth labels with the predictions from the models trained during the latest iteration. In this situation, if the model performs badly in one iteration, the misclassified labels are carried forward in further iterations, thus affecting the training adversely. 

To address this issue, we modify Algorithm \ref{alg:iter_train} to now maintain the predictions of the best model obtained so far ($S_{best}$), and greedily select the labels of a fraction of the samples from the corresponding predictions in $S_{best}$, while the remaining sample labels are obtained from the predictions of the latest trained model. These labels serve as pseudo-ground truths for iterative training. This allows us to partially leverage the predictions of the best model so far, and is hence termed as \emph{Semi-Greedy Iterative Training (SGIT)} approach.

The model is trained for $T$ iterations where in each iteration $t + 1$, we greedily sample $\beta \in (0, \; 1)$ fraction of triplets and their corresponding predictions from the best model obtained so far, while the labels for the remaining $(1 - \beta)$ fraction of triplets are obtained from the predictions of the model at iteration $t$. Note that $\beta$ decides the reliance of our technique on the best model obtained so far. Setting $\beta = 0$ would be effectively the same as Algorithm \ref{alg:iter_train}, while setting $\beta = 1$ would stop the \emph{exploration} process once a \emph{best} model is found, although the accuracy of that best model may not be up to the mark. Inspired from the concept of multi-armed bandits, where sublinear regret is achieved only if there is some infinite amount of exploration, we claim that setting $\beta = 1$ would not allow us to reach the best model possible. 

Now, as indicated in Section \ref{expt2}, $\alpha \in (0, \; 1)$ fraction of triplets from this augmented dataset are used to train the model for iteration $t + 1$. The detailed method is described in Algorithm \ref{alg:semi_greedy_iter_train}.

\begin{algorithm}
\SetAlgoLined
     \SetKwInOut{Input}{input}
     \SetKwInOut{Output}{output}
     \Input{samples $S = \{(x_i, \; y_i)\}_{i=1}^n$, number of iterations $T$, fraction of samples $\alpha$, greediness factor $\beta$}
     
     $\theta_0 \gets$ pretrained \texttt{BERT} embeddings $+$ random weights\;
     $\theta^* \gets \theta_0$\;
     $acc_{best} \gets accuracy(\{y_i\}_{i=1}^n, \;  \{\theta_0(x_i)\}_{i=1}^n)$\;
     $S_{best} \gets S$\;
     \For{t = 0, 1, ..., T-1}{
        $S_{t_{best}} \stackrel{i.i.d}{\sim} Uniform(S) \;$ s.t. $| \; S_t \; | = \lfloor \beta n \rfloor $\;
        $S_{t_{latest}} \gets \{(x_j, \; y_j)\} \; | \; (x_j, *) \in S \; \land \; (x_j, *) \notin S_{t_{best}}$\;
        \tcc{$| \; S_{t_{latest}} \; | = \lceil (1 - \beta) n\rceil$}
        \
        
        $S_{temp} = S_{t_{best}} \cup S_{t_{latest}}$ \tcp*{$| \; S_{temp} \; | = n$}
        $S_t \stackrel{i.i.d}{\sim} Uniform(S_{temp}) \;$ s.t. $| \; S_t \; | = \lfloor \alpha n \rfloor $\;
        $\theta_{t+1} \gets TrainClassifier(\theta_t, S_t)$\;
        \
        
        $pred \gets \{\theta_{t+1}(x_i)\}_{i=1}^n$\;
        $acc_{curr} \gets accuracy(\{y_i\}_{i=1}^n, \;  pred)$\;
        \If{$acc_{curr} > acc_{best}$}{
            $acc_{best} \gets acc_{curr}$\;
            $\theta^* \gets \theta_{t+1}$\;
        }
        
        $S \gets \{(x_i, \; pred_i)\}_{i=1}^n$\;
        
     }
     \Output{$\theta^*$}
     
    \caption{Semi-Greedy Iterative Training (\texttt{SGIT})}
    \label{alg:semi_greedy_iter_train}
\end{algorithm}

\begin{figure}[!h]
  \centering
  \includegraphics[width=\linewidth]{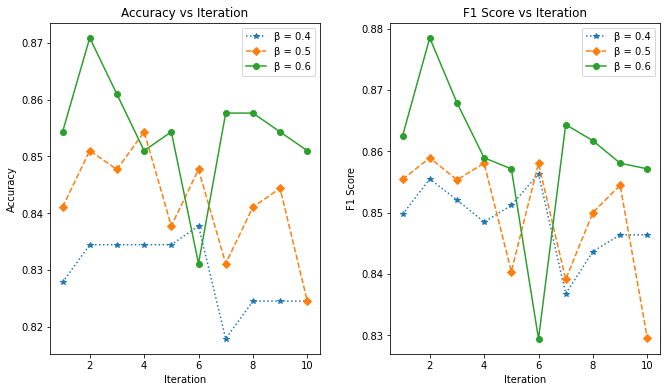}
  \caption{Accuracy and F1 scores of the models for}
  \label{fig:expt_4_metrics}
\end{figure}

We set $\alpha = 0.8$ and study the impact of changing $\beta$ on the performance of our transformer-based model. For this we train the model for $T = 10$ iterations on the dataset containing $64000$ triplets, using $\beta \in \{0.4, 0.5, 0.6\}$. The performances of different models against our test set are plotted in Figure \ref{fig:expt_4_metrics}. The best model performance is obtained in iteration 2 for $\beta = 0.6$, where we achieve an accuracy of $87.08 \%$ and F1 score of $0.8785$. It is observed that in general, increasing $\beta$ helps in obtaining a better classifier. This could potentially be explained by improvement in the pseudo-ground truths when one continues using more labels from the best model instead of relying heavily only on the latest iteration. Analysing the confusion matrices suggests that the best model trained using $\beta = 0.6$ improves its accuracy and F1 score by correcting most of the exploratory intent queries like \emph{`mobile basics'} and \emph{`english kids'}, which were misclassified by the best model trained with $\beta = 0.4$.

\vspace{-10pt}

\subsection{Testing the Model on Tail Queries} \label{expt5}

Based on the popularity of the queries, queries are typically divided into three categories: \emph{Head}, \emph{Body} and \emph{Tail}. As explained in Section \ref{data_prep_bipartite}, a query must have at least $40$ related queries in the set $Q_n$ for our heuristic. By definition, such queries belong to the Head or Body segments. However, a significantly large fraction of unique user queries belong to the Tail segment in the search log. Such queries have much smaller number of related queries. In this experiment, we test the performance of our model on such tail queries. Towards that, we uniformly sample a set of $96$ queries, which have 1 to 9 related queries. These queries are judged by the human judges, following the same methodology, as described in Section \ref{judging}.  Next, we label these queries based on our best performing model (cf. Section \ref{expt4}). Results for this experiment are presented in Table \ref{tab:expt_5_tail_queries_cf}. We notice that our model achieves an accuracy of $74.73 \%$, and F1 score of $0.7736$, denoting that it generalizes well over different segments of queries. Examples of tail queries that were classified correctly by our transformer-base model include \emph{`casa grande broad street richmond va'} (lookup) and \emph{`brownside gang'} (exploratory).

\begin{table}[!h]
    \caption{Confusion Matrix for tail queries classification}
    \label{tab:expt_5_tail_queries_cf}
    \begin{tabular}{l|l|c|c|}
    \multicolumn{2}{c}{}&\multicolumn{2}{c}{\textbf{Predicted Label}}\\
    \cline{3-4}
    \multicolumn{2}{c|}{}& \textbf{Exploratory} & \textbf{Lookup}\\
    \cline{2-4}
    \multirow{2}{*}{\textbf{True Label}}& \textbf{Exploratory} & $31.58 \%$ & $10.53 \%$\\
    \cline{2-4}
    & \textbf{Lookup} & $14.74 \%$ & $43.16 \%$\\
    \cline{2-4}
    \end{tabular}
\end{table}

\section{Analysis and Discussion} \label{analysis_discussion}

Existing methodologies to classify the queries into exploratory or lookup revolve around designing well-defined tasks and studying the user behavior on such tasks \cite{Herrera10, Zhang19, Aula10, Athukorala14}. They analyze parameters such as query length, query duration, scroll depth, cumulative clicks, task completion time, dwelling duration, etc., to conclude the specificity of a query. In these experiments, the final query set obtained is not beyond a few thousand queries, marked as exploratory or lookup, and fail to capture the generic and diversified query intents expressed in millions of user queries. We develop a novel technique that scales to a large number of user search queries, well distributed over multiple user intents.

The accuracy of our heuristic is found to be 81.12\%, based on human judgements over a randomly sampled query set. Therefore, the statistical trends for the queries, labelled by our heuristic, can be assumed to be applicable on the entire class of queries. As shown in the various experiments above, analysis of such queries at scale showed that the factors such as query length, session length, and the query position in the session play little role in identifying the query specificity, contrary to the existing accepted intuitions. 

We further showed that the performance of our transformer-based model outperforms the high bar of our heuristic. Our best performing model had an accuracy of 87.41\%. The prediction time of our model makes it a promising candidate to be incorporated into search engines to improve search results at runtime.

\section{Conclusion and Future Work} \label{conclusion}

In this paper, we presented a novel methodology to identify the user query specificity that is scalable to highly diversified web queries. Our first contribution is a heuristic-based methodology to label the queries according to their specificity. We showed that on the human-judged samples, our heuristic performed well. To the best of our knowledge, ours is the first method to scale to such a large number of queries and captures the diverse intents in web search. Further, ours is the first study to show that the factors such as query length and a few of the session characteristics, hitherto held important, do not play a very important role in defining the query specificity, when this data was analyzed statistically at scale. 

We further proposed a transformer-based model over the data labelled by our unsupervised heuristic, which is considered the pseudo ground truth. We proposed a novel training methodology, SGIT (Semi-Greedy Iterative Training), and showed that our model outperformed the heuristic baseline on the human judged query set. We further showed that the model performance improves as the training sample size is increased, signifying the generalization capabilities of our model as well as quality of the underlying training data labelled using our heuristic.

As part of our future work, we plan to further analyze the statistical insights of the queries with different query specificities. Such insights may have significant impact on analyzing the performance of search engines. We further plan to work on optimizing our deep learning model. We also plan to publicly release the sampled data to enable further study over such queries at scale.

\balance
\bibliographystyle{ACM-Reference-Format}
\bibliography{references}

\end{document}